\documentclass[twocolumn,amsmath,amssymb,prl]{revtex4}
\usepackage{graphicx}

\begin{document}

\title{Structural and electronic phase transitions driven by electric field in metastable MoS$_2$ thin flake}

\author{C. Shang$^{1 }$\footnotemark[2], B. Lei$^{1 }$\footnotemark[2], W. Z. Zhuo$^1$, Q. Zhang$^1$, C. S. Zhu$^1$, J. H. Cui$^1$, X. G. Luo$^1$, N. Z. Wang$^1$, F. B. Meng$^1$, L. K. Ma$^1$, C. G. Zeng$^1$, T. Wu$^{1,3}$, Z. Sun$^{2,3}$, F. Q. Huang$^{3,4,5}$,}
\author{X. H. Chen$^{1,6,7}$}
\email{chenxh@ustc.edu.cn}

\affiliation{
$^1$Key Laboratory of Strongly-coupled Quantum Matter Physics, Chinese Academy of Sciences, and Hefei National Laboratory for Physical Sciences at Microscale, and Department of Physics, University of Science and Technology of China, Hefei, Anhui 230026, China\\
$^2$National Synchrotron Radiation Laboratory, University of Science and Technology of China, Hefei, Anhui 230026, China\\
$^3$CAS Center for Excellence in Superconducting Electronics (CENSE), Shanghai 200050, China\\
$^4$State Key Laboratory of High Performance Ceramics and Superfine Microstructure, Shanghai Institute of Ceramics, Chinese Academy of Sciences, Shanghai 200050, China\\
$^5$State Key Laboratory of Rare Earth Materials Chemistry and Applications, College of Chemistry and Molecular Engineering, Peking University, Peking 100871, China\\
$^6$CAS Center for Excellence in Quantum Information and Quantum Physics, Hefei, Anhui 230026, China\\
$^7$Collaborative Innovation Center of Advanced Microstructures, Nanjing University, Nanjing 210093, China}

\date{\today}

\begin{abstract}
Transition-metal-dichalcogenides own a variety of structures as well as electronic properties which can be modulated by structural variations, element substitutions, ion or molecule intercalations, etc. However, there is very limited knowledge on metastable phases of this family, especially the precise regulation of structural changes and accompanied evolution of electronic properties. Here, based on a new developed field-effect transistor with solid ion conductor as the gate dielectric, we report a controllable structural and electronic phase transitions in metastable MoS$_2$ thin flakes driven by electric field. We found that the metastable structure of 1T$^{'''}$-MoS$_2$ thin flake can be transformed into another metastable structure of 1T$^{'}$ -type upon intercalation of lithium regulated by electric field. Moreover, the metastable 1T$^{'}$ phase persists during the cycle of intercalation and de-intercalation of lithium controlled by electric field, and the electronic properties can be reversibly manipulated with a remarkable change of resistance by four orders of magnitude from the insulating 1T$^{'}$-LiMoS$_2$ to superconducting 1T$^{'}$-MoS$_2$. Such reversible and dramatic changes in electronic properties provide intriguing opportunities for development of novel nano-devices with highly tunable characteristics under electric field.
\end{abstract}



\maketitle

\footnotetext[2]{These authors contributed equally to this work.}

A variety of novel physical properties are accommodated in transition-metal-dichalcogenides (TMDs) owing to the versatile composition, structure and dimensionality as well as the concomitant electronic structures \cite{1,2,3,4,5,6,7,8}. Intensive studies of TMDs demonstrate that these two-dimensional materials have promising application in modern science and technology, such as optoelectronics \cite{4,6}, valleytronics \cite{5,9}, heterostructure transistor \cite{10,11,12,13}, etc. By controlling carrier concentration \cite{14,15,16} and ion or molecule intercalations \cite{17,18,19}, the tunability of structural characteristics and electronic properties can be further extended. Many TMDs with stable 1T, 2H and 3R lattice structures have been well studied. However, there exists various metastable TMDs with intriguing properties, which demand elaborate investigations. For instance, three metastable polytypes have been observed in MoS$_2$, with different superlattices, namely, a$\times$2a (1T$^{'}$), 2a$\times$2a (1T$^{''}$) and $\sqrt{3}$a$\times$$\sqrt{3}$a (1T$^{'''}$) \cite{20} (see Fig. 1, which are energetically less stable than 2H structure \cite{22,23}.  Theoretical studies suggest that these metastable polytypes may host exotic physical properties, such as ferroelectricity \cite{22} and quantum spin Hall effect \cite{21}. In particular, how to effectively regulate the electronic behavior of these metastable polytypes needs more careful studies, which would promote the invention and application of new nano-devices based on these materials.

 In TMD materials, the ground state energy of various polytypes are very different \cite{23}. The lattice distortion develops upon introducing extra electrons by intercalated alkali metal, and various metastable superstructures can be induced after the removal of extra electrons during oxidation process. This is because the extra electrons at high doping level can raise the total energy of a specific system, which overcomes the transition barrier between different lattice structures and makes structural transitions possible. Locally, the extra electrons cause a change in the electron count of \textit{d} orbitals, leading to the instability of the original lattice structure and resulting in a new metal coordination. In light of this mechanism, electrostatic doping based on FET (field-effect transistor) device has been applied to monolayer MoTe$_2$ to realize a reversible transformation between 2H phase and 1T$^{'}$ phase \cite{14}. Traditional FET technology can inject carrier into a thin channel via electrostatic doping, so it can be utilized to modulate lattice structures of few-layer TMDs to reach the metastable lattice distortions. However, such FET configuration performs effective charge doping only in thickness of few nanometers and electrons cannot be transported to deeper regime due to Thomas-Fermi screening \cite{30,31}, which is thus not suitable for tuning electronic properties and metastable polytypes in a bulk crystal with thickness more than tens of nanometers.

 Recently, we have developed a new type of FET configuration using solid ion conductor (SIC) as gate dielectric.  \cite{32,33,34,35}. Taking advantage of the relatively long migration length of lithium ions, we can dope carriers uniformly into thin flakes with thickness of tens of nanometers, largely surpassing the shallow doping regime in the electrostatic-type charge doping and modulating the bulk. In this way, the intercalation and de-intercalation of ions between the constituent layers can modulate the electronic properties and lattice structures. In addition, the implantation of external ions can modulate the structure of the parent material directly to some extent, transcending the influence caused merely by electrostatic doping. This paves a way for exploring more abundant metastable structures and exotic electronic properties that could be hosted in TMD materials.

 \begin{figure}[h]
 	\centering
 	\includegraphics[width=0.45\textwidth]{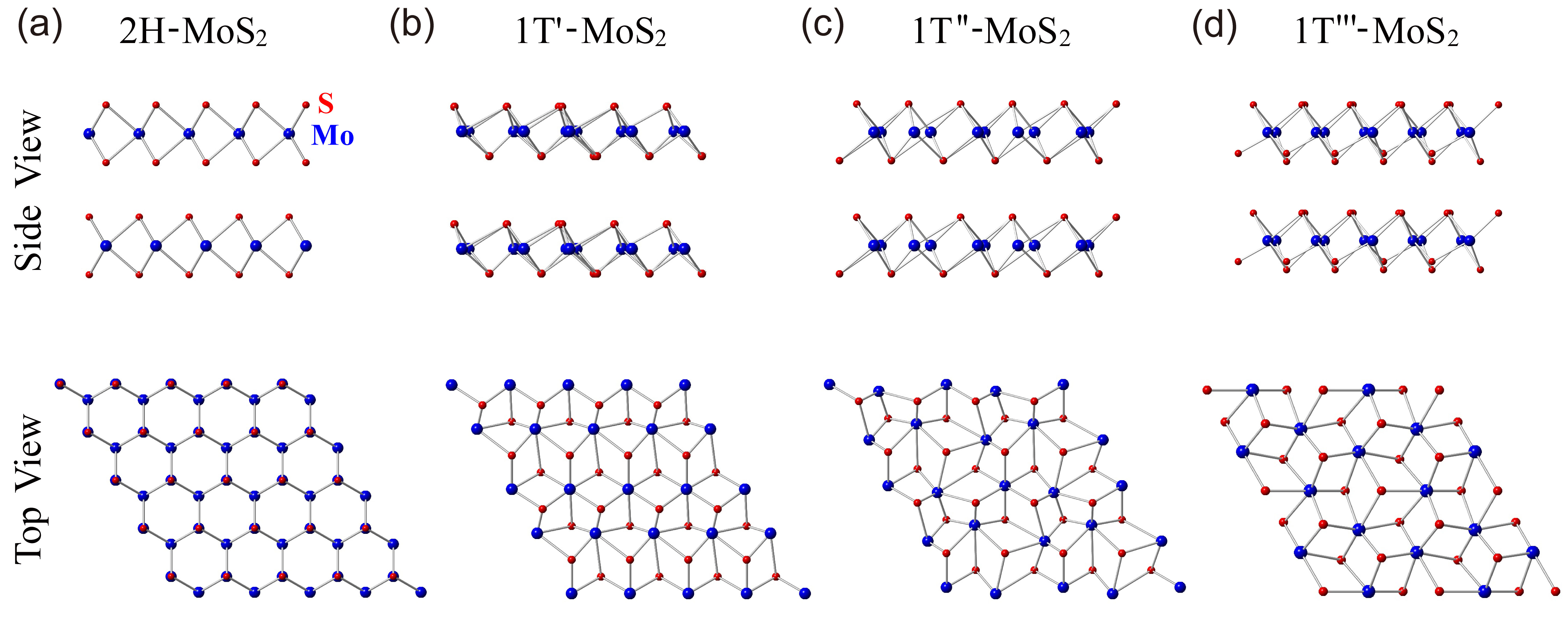}
 	\caption{(color online) Lattice structures of stable 2H phase and metastable 1T-type polytypes of MoS$_2$. (\textbf{a})-(\textbf{d}): The side views (top row) and top views (bottom row) of 2H-MoS$_2$, 1T$^{'}$-MoS$_2$, 1T$^{''}$-MoS$_2$ and 1T$^{'''}$-MoS$_2$, respectively. The atoms of Mo and S are labelled and displayed with different balls.}
 	\label{Fig. 1}
 \end{figure}

 In this letter, we demonstrate the regulation of metastable MoS$_2$ polytypes using Li ions with SIC-FET device. We start with 1T$^{'''}$-MoS$_2$ thin flake, and upon applying gate voltage Li intercalation leads to the development of insulating LiMoS$_2$ with 1T$^{'}$ polytype of MoS$_2$ layers.  When lithium ions are expelled from the thin flake with the variation of gate voltage, 1T$^{'}$ distortion persists and the thin flake turns into 1T$^{'}$-MoS$_2$ superconductor. The transition between the insulating 1T$^{'}$-LiMoS$_2$ and the 1T$^{'}$-MoS$_2$ superconductor is reversible by controlling the intercalation and de-intercalation of lithium ions with gate voltage. As the gate voltage changes, resistance measurements show a hysteretic loop, indicating that the transition between the superconductor and insulator is similar to the typical characteristic associated with the first-order phase transition. These structural transitions can be well distinguished by Raman spectroscopy.

We have recently synthesized high-quality crystals of metastable 1T$^{'''}$- and 1T$^{'}$-MoS$_2$ and studied their properties  (see the supplementary materials and Ref. \cite{26,24} for more details), which can serve as a reference to characterize the thin flake at different stages in our SIC-FET device . Fig. 2(a) illustrates the atomic-resolution STM image of 1T$^{'''}$-MoS$_2$ insulator, and the periodicity of $\sqrt{3}$a $\times$ $\sqrt{3}$a is evidence of 1T$^{'''}$ type lattice distortion. The dI/dV spectrum in Fig. 2(b) shows a gap of 0.6 eV in this insulator, which is in agreement with the gap size predicted by band theory calculations \cite{22}. Fig. 2(c) depicts the configuration of the device used to regulate lattice structures as well as the associated electronic behavior. Fig. 2(d) shows the initial 1T$^{'''}$-MoS$_2$ thin flake with Hall bar electrodes attached. In our experiments, the size of the sample is $\sim$ 30 microns and the thickness is 80 nm (see Fig. 2(e)). Fig. 2(f) is a typical evolution of resistance upon lithium intercalation and de-intercalation driven by electric field, demonstrating drastic changes in electronic properties. The gate voltage was applied with a scan rate of 1 mVs$^{-1}$ at $T$ = 260 K, at which temperature the migration of lithium can be well controlled by gate voltage.

\begin{figure}[h]
\centering
\includegraphics[width=0.45\textwidth]{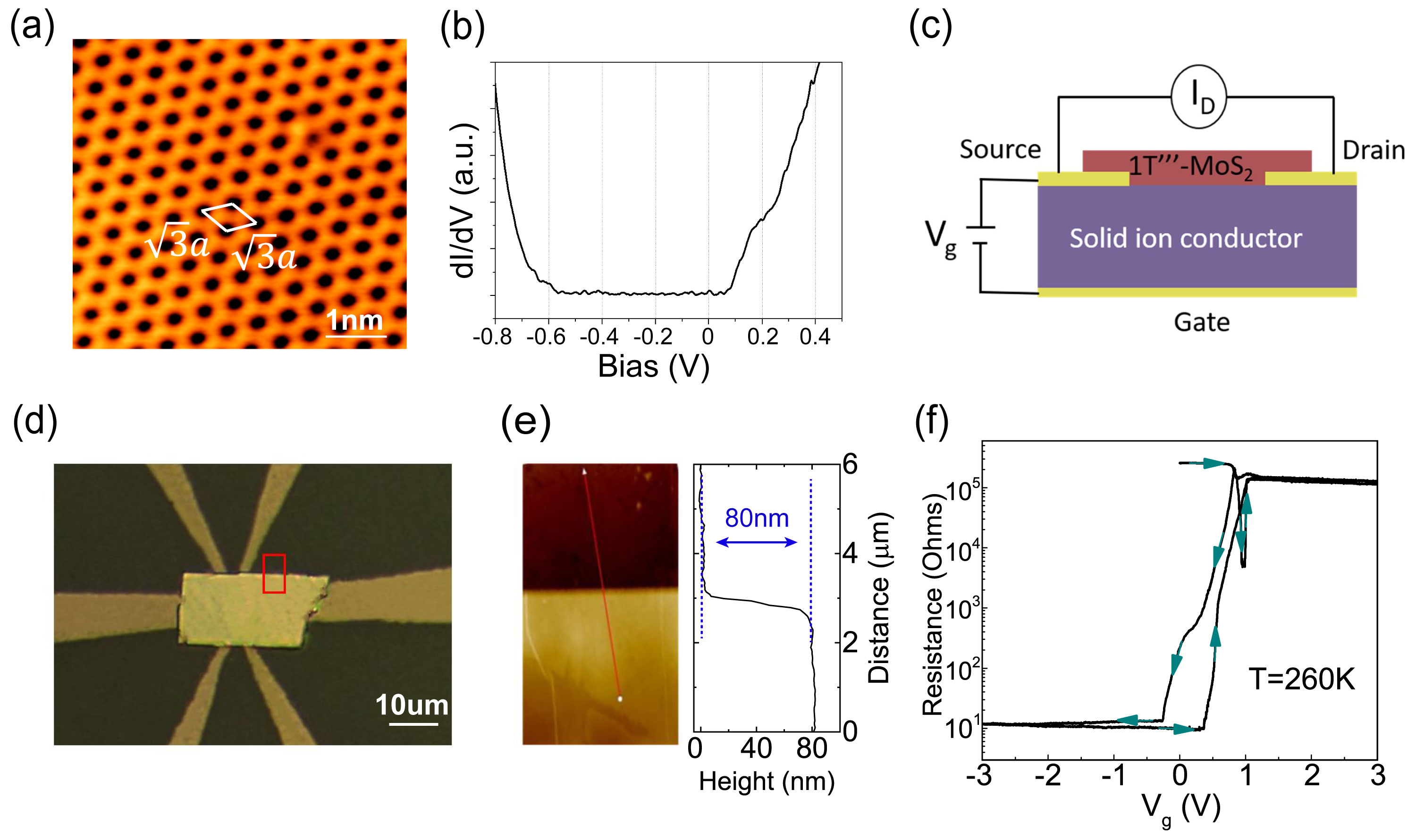}
\caption{(color online) The modulation of metastable MoS$_2$ by gate voltage with SIC-FET device. (\textbf{a}): Atomic-resolution STM image (V$_s$ = 6.2 V, I$_t$ = 57 pA) of 1T$^{'''}$-MoS$_2$ insulator. The periodicity of $\sqrt{3}$a $\times$ $\sqrt{3}$a (a = 0.31 nm) is marked. (\textbf{b}): The dI/dV spectrum shows a gap of $\sim$ 0.6 eV for 1T$^{'''}$-MoS$_2$ insulator. (\textbf{c}): A schematic view of the SIC-FET device fabricated with MoS$_2$ thin flake. From the bottom to the top: Cr/Au back gate layer, finely polished Li ion conductor substrate, MoS$_2$ thin flake and Cr/Au Hall bar electrodes. (\textbf{d}): The optical image of a 1T$^{'''}$-MoS$_2$ thin flake with a standard Hall bar configuration, and the current and voltage terminals are labeled. (\textbf{e}): The thickness of 1T$^{'''}$-MoS$_2$ thin flake. The image (left) is an enlarged view of the portion enclosed by the red rectangle. The profile (right) from atomic force microscopy shows the flake thickness of 80 nm. (\textbf{f}): Gate voltage dependent resistance of a MoS$_2$ thin flake with thickness of $\sim$ 80 nm in SIC-FET device. The continuously swept gate voltage is applied at $T$ = 260 K with a scan rate of 1 mVs$^{-1}$.}
\label{Fig. 2}
\end{figure}

We can divide the whole gating process into two stages with irreversible and reversible characteristics, respectively, as shown in Figs. 3(a) and 3(b). With the intercalation of lithium ions, the transition from 1T$^{'''}$-MoS$_2$ (Phase I) to 1T$^{'}$-LiMoS$_2$ (Phase II) is irreversible (Fig. 3(a)), which is further confirmed by Raman measurements (Fig. 4(d)). More details on 1T$^{'}$-LiMoS$_2$ will be discussed later. The dip in resistance indicates the range of bias voltage in which the structural transition takes place. Subsequently, a further regulation by gate voltage can drive the intercalation and de-intercalation of lithium ions, leading to a reversible transition between the 1T$^{'}$-LiMoS$_2$ and 1T$^{'}$-MoS$_2$ (Fig. 3(b)). In order to determine the electronic properties at different gating status, we measured temperature-dependent resistances. As shown in Fig. 3(c), the thin flake is insulating at $V_g$ = 0 V, 0.95 V and 3 V, though the electronic behavior is evidently different. At $V_g$ = -3 V, the resistance of the thin flake is lowered by four orders of magnitude with respect to $V_g$ = 3 V, showing a metallic behavior with a superconducting transition at $T_c$ = 4.2 K. In combination of Raman spectra (as will be shown later) and resistance taken at different stages, we identify that,  at $V_g$ = 0 V, 3 V, -3 V, the thin flake are 1T$^{'''}$-MoS$_2$, 1T$^{'}$-LiMoS$_2$ and 1T$^{'}$-MoS$_2$, respectively. The insulating behavior of 1T$^{'''}$-MoS$_2$ and 1T$^{'}$-LiMoS$_2$ is consistent with theoretical calculations \cite{22,36,37}. At $V_g$ = -3 V, the superconducting transition temperature and Raman modes are identical to those of 1T$^{'}$-MoS$_2$ as we reported in Ref. \cite{26,24}.

\begin{figure}[h]
	\centering
	\includegraphics[width=0.45\textwidth]{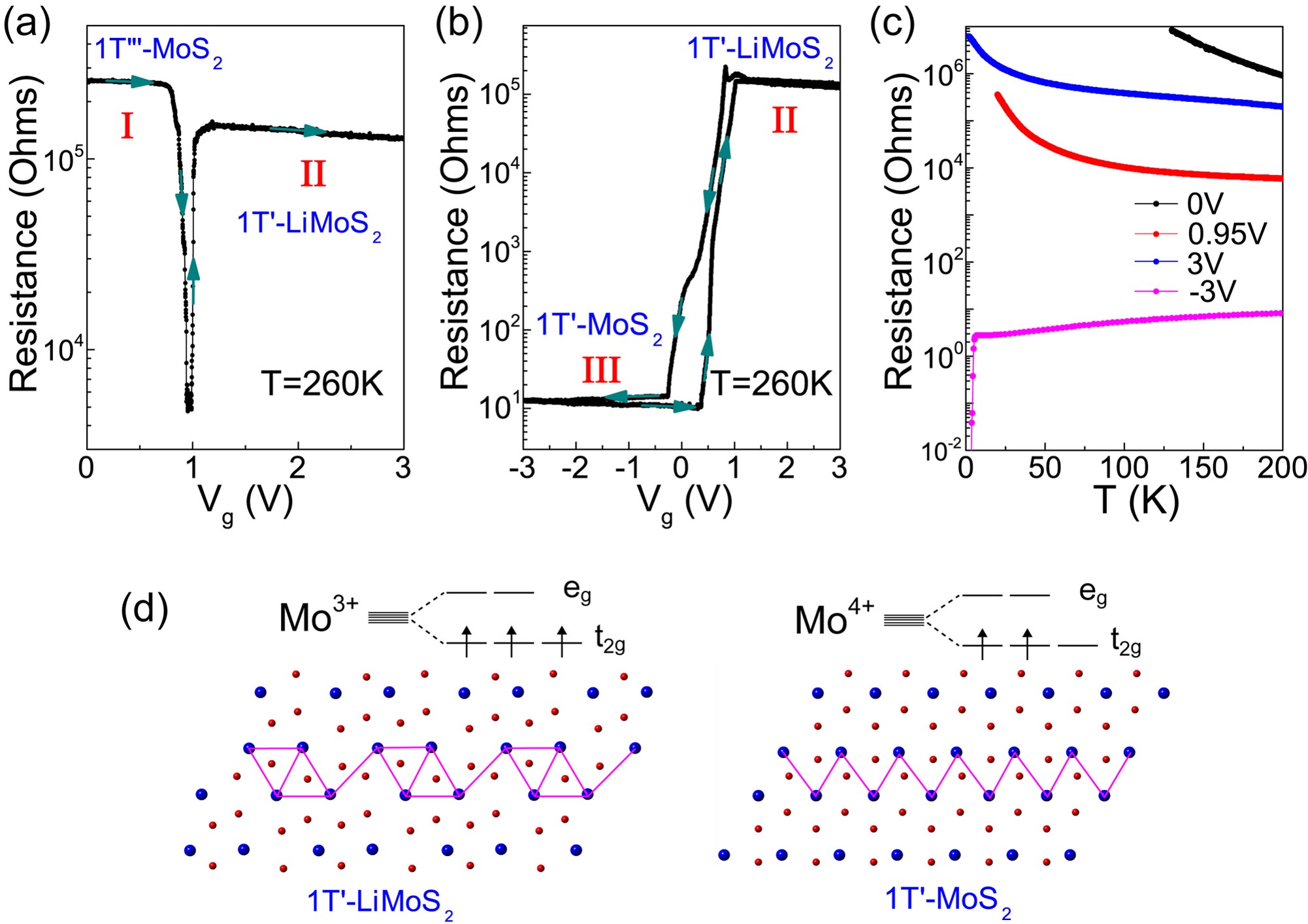}
	\caption{(color online) Resistance of MoS$_2$ thin flake as a function of gate voltage and temperature. (\textbf{a}): Gate voltage dependent resistance of a MoS$_2$ thin flake in the irreversible transition. (\textbf{b}): Gate voltage dependent resistance in the reversible transition. (\textbf{c}): The resistance measured at $V_g$ = 0, 0.95, 3 and -3 V. For $V_g$ = 0, 0.95 and 3 V, the thin flake is in different insulating states, while it becomes metallic and superconducting at $V_g$ = -3 V. (\textbf{d}): A schematic picture of local electron counts and Mo-Mo bonds in  1T$^{'}$-LiMoS$_2$ and  1T$^{'}$-MoS$_2$ }
	\label{Fig. 3}
\end{figure}

In these materials, the main physics occurs in MoS$_2$ layers, and the primary role of Li ions located between the layers is to provide electron carriers. Theoretical calculations and experimental investigations suggest that, with the intercalation of Li, 1T$^{'}$-LiMoS$_2$ emerges \cite{36,37,38}. In LiMoS$_2$, Mo ions are in the 3+ state, with 3 \textit{d}-electrons in an octahedral coordination. Locally, it has unoccupied $e_g$ orbitals and half-filled triply degenerate $t_{2g}$ orbitals, as shown in Fig. 3(d). Each electron in $t_{2g}$ orbital pairs with that of a neighboring Mo ion. Therefore, each Mo have three Mo-Mo bonds and forms diamond-like chains. In both of 1T$^{'''}$- and 1T$^{'}$-MoS$_2$, Mo ions are in 4+ state with 2 $t_{2g}$ electrons. The intercalation of Li introduces extra electrons into MoS$_2$ layers and give rise to 3+ Mo ions, which eventually trigger the rearrangement of Mo-Mo bonds to lower the total energy, and then leads to the particular atomic configurations in 1T$^{'}$-LiMoS$_2$. With this type of atomic configurations, 1T$^{'}$-LiMoS$_2$ is an insulator as indicated by theoretical calculations  \cite{36,37}.

As shown in Fig. 3(a), with increasing the gate voltage, the resistance of the thin flake first drops nearly by about two orders of magnitude at $V_g$ = 0.95 V, in accordance with slight electron doping by Li intercalation in the insulating 1T$^{'''}$-MoS$_2$. When LiMoS$_2$ grows with more Li intercalation, the rise of resistance suggests that the insulating behavior becomes stronger again. During this process, individual MoS$_2$ layers changes from 1T$^{'''}$ to 1T$^{'}$ configuration. With further increasing the gate voltage, the resistance remains almost at the same level, suggesting that the transition from 1T$^{'''}$-MoS$_2$ to 1T$^{'}$-LiMoS$_2$ is finished when $V_g$ is above 1.2 V. When the gate voltage is swept backwards, the lithium de-intercalation does not induce the structural change back to 1T$^{'''}$-MoS$_2$. Instead, the structure and electronic properties enter a reversible cycle, as shown in Fig. 3(b).

In Fig. 3(b), when the gate voltage sweeps gradually from 3 V to lower values, the resistance remains almost the same. Then, the resistance begins to drop drastically at $V_g$ = 0.7 V, suggesting that the entire sample begins to evolve from an insulator to a metal during the lithium de-intercalation process. When the gate voltage reaches $V_g$ = -0.25 V, the resistance already falls by four orders of magnitude. With the further enhancement of the negative gate voltage, the resistance nearly keeps constant, indicating that the lithium is almost completely driven out of the thin flake and the sample becomes 1T$^{'}$-MoS$_2$. Based on the atomic configuration pictures as presented in Fig. 3(d), the transition from insulator to metal occurs in the following way. With the de-intercalation of Li, Mo atoms changes from Mo$^{3+}$ in 1T$^{'}$-LiMoS$_2$ to Mo$^{4+}$ in 1T$^{'}$-MoS$_2$, and there are two \textit{d}-electrons left in the $t_{2g}$ orbitals, which form Mo-Mo bonds with two neighboring Mo ions, as the zigzag chains illustrated in the schematic plot of 1T$^{'}$-MoS$_2$. The different occupation of $t_{2g}$ orbitals indicates that 1T$^{'}$-LiMoS$_2$ is an insulator, and 1T$^{'}$-MoS$_2$ is a metal.  From this perspective, this insulator-metal transition is due to electronic effect with accompanied structural changes. Theoretical calculations of 1T$^{'}$-LiMoS$_2$ compound show that the overall distribution of density of states are not affected by the formation of Mo-Mo bonds \cite{36}. This suggests that there is no significant electron correlation or localization for the drastic electronic and structural changes.

When the gate voltage is reversed towards positive values, the resistance barely varies from $V_g$ = -3 V to 0.4 V. Then the resistance increases sharply, indicating that the intercalation of lithium drives the transition towards the insulating 1T$^{'}$-LiMoS$_2$. When the gate voltage reaches $V_g$ = 1 V again, the resistance is accumulatively increased by four orders of magnitude, and the sample enters the 1T$^{'}$-LiMoS$_2$ phase. It is noteworthy that this $V_g$ value coincidences with the gate voltage value of the transition from 1T$^{'''}$-MoS$_2$ to 1T$^{'}$-LiMoS$_2$ in the irreversible process shown in Fig. 3(a). With the further increase of gate voltage, the resistance remains basically unaffected. Our studies suggest that the transition between 1T$^{'}$-MoS$_2$ and 1T$^{'}$-LiMoS$_2$ is reversible.

Since phonon modes vary significantly in different lattice configurations, we use Raman spectroscopy to identify the evolution of lattice structures with the gate voltage. Initially, the thin flake is in the 1T$^{'''}$ phase, and its Raman spectrum is shown in Fig. 4(a), which is consistent with our previous experimental reports and theoretical simulations \cite{26}. At $V_g$ = 3 V, the intercalation of lithium causes significant changes in the Raman spectra as shown in Fig. 4(b). It is highly consistent with the Raman spectra of Li-intercalated 1T$^{'}$ polytype of MoS$_2$ \cite{39,40,41}, which can serve as a reference to prove that the lattice configuration of MoS$_2$ layers has changed to 1T$^{'}$ structure upon Li intercalation. At $V_g$ = -3 V, the lithium ions have been removed from the sample, and the Raman spectra in Fig. 4(c) is in excellent agreement with the standard 1T$^{'}$-MoS$_2$ Raman spectra \cite{26,24,27}, indicating that the sample is 1T$^{'}$-MoS$_2$. Comparing to the pristine 1T$^{'}$-MoS$_2$, the relative changes in the peak intensity could be due to the defects caused by the migration of Li ions. In particular, the resistance measurements also indicate that the metallicity and superconductivity are highly consistent with previous reports on 1T$^{'}$-MoS$_2$ superconductor \cite{24,25,26}. We have also performed high spatial-resolution Raman measurements to study the structural homogeneity of the sample at $V_g$ = 0 V, 3 V and -3 V, and found that the structural transitions are uniform in the thin flake upon the intercalation and de-intercalation of Li ions (see Fig. S5 in supplementary materials). Using Raman mode as a fingerprint, we demonstrate in Fig. 4(d) how MoS$_2$ layers evolve from 1T$^{'''}$-MoS$_2$ to 1T$^{'}$-LiMoS$_2$. When a small amount of Li ions is driven into the 1T$^{'''}$ thin flake, there is no evident for Raman modes of 1T$^{'}$ phase. As the intercalated Li ions increase, the 1T$^{'}$ configuration emerges and its percentage grows in line with the drastic increase of resistance, while the 1T$^{'''}$ component decreases gradually and eventually vanish when the resistance reaches the saturated value.

\begin{figure}[h]
\centering
\includegraphics[width=0.45\textwidth]{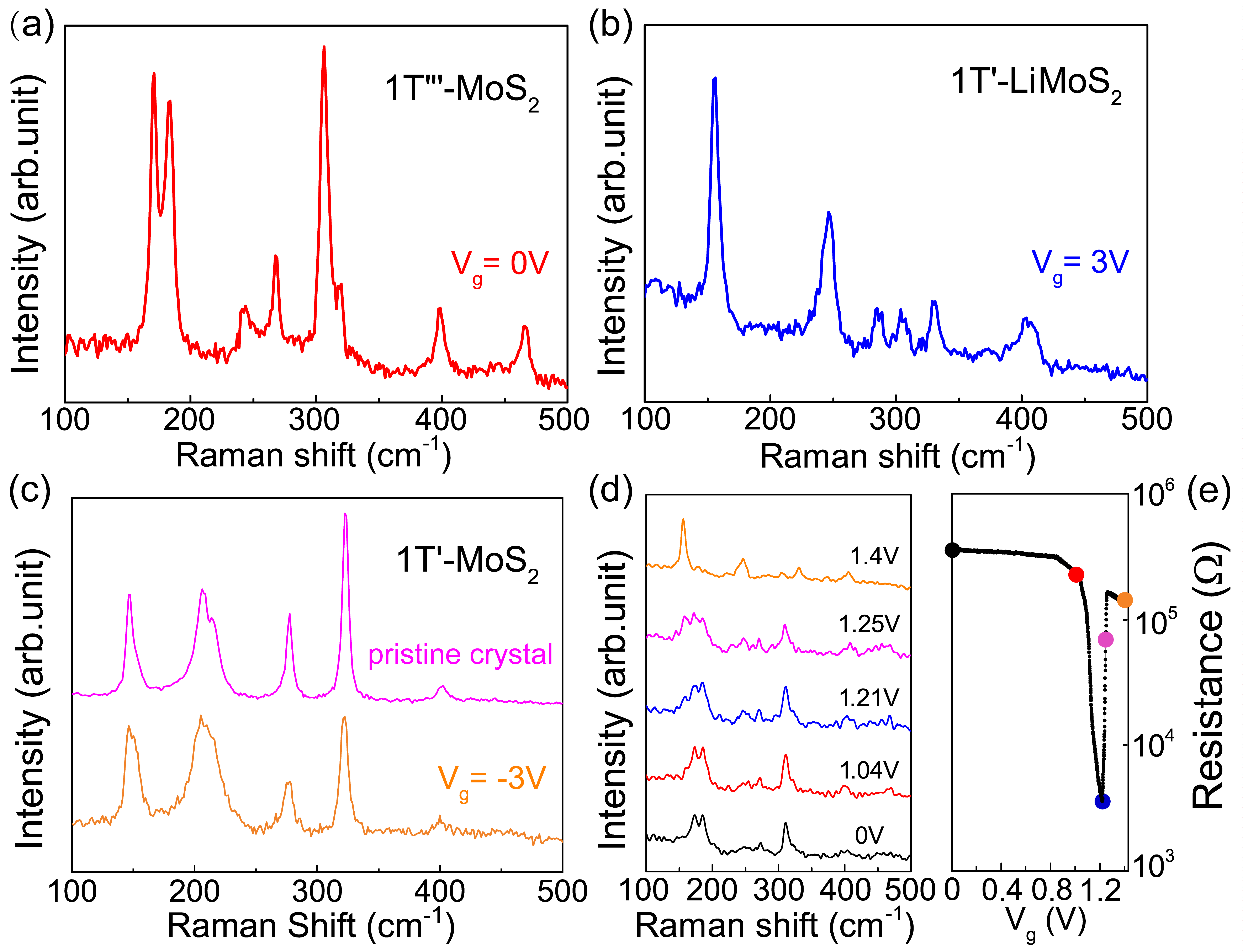}
\caption{(color online) Raman spectra accompanying structural transitions induced by gating. (\textbf{a}): Raman modes taken from pristine 1T$^{'''}$-MoS$_2$ thin flake. (\textbf{b}): Raman spectrum for the MoS$_2$ thin flake at $V_g$ = 3 V. (\textbf{c}): Raman spectrum taken from a pristine 1T$^{'}$-MoS$_2$ crystal and the MoS$_2$ thin flake at $V_g$ = -3 V. (\textbf{d}): Raman spectra taken under various gate voltages. It shows how the lattice structure evolve from 1T$^{'''}$ to 1T$^{'}$ configurations. (\textbf{e}): The corresponding resistance for different Raman spectra shown in panel d.}
\label{Fig. 4}
\end{figure}

During the reversible transition controlled by the SIC-FET device, the main change comes from the intercalation and de-intercalation of lithium ions, while the 1T$^{'}$  lattice distortion of MoS$_2$ layers remains with some variation of Mo-Mo bonds. With the insertion of Li ions, extra electrons provided by Li can induce the development of 1T$^{'}$-LiMoS$_2$ with three neighboring Mo-Mo bonds, which has a band gap at the Fermi level and shows insulating behavior. When Li ions are driven out of MoS$_2$ lattice, 1T$^{'}$-MoS$_2$ forms with two neighboring Mo-Mo bonds, exhibiting metallic properties. When the applied gate voltage reaches a certain threshold, the lithium intercalation will occur and 1T$^{'}$-LiMoS$_2$ will develop; when the gate voltage is reduced to a certain level, the lithium ion will be de-intercalated, and at a certain negative bias, most lithium ions will be driven out of the 1T$^{'}$-MoS$_2$ lattice. Therefore, the overall transition between the metal and the insulator looks similar to the hysteresis loop of a first-order phase transition.

The gating technique used in the present study is significantly different from the traditional FET technology. In the traditional approach, electronic properties can only be modulated in few layers near the interface of the sample and dielectric \cite{30,31}. Taking advantage of the long migration length of lithium ions, our method can change the lattice structures and electronic properties in the thickness of tens of nanometers, and yields more bulk-like modulations. Especially, the introduction of external ions provides a wider regulatory range for the structural and electronic characteristics. Based on the present findings, such all-solid-state devices is of great potential for electronic switches, memory switches, reconfigurable circuits, and so on.

In summary, we have shown a remarkable transition of structure and electronic properties in 1T$^{'''}$- and 1T$^{'}$-MoS$_2$ lattice, which can be induced by lithium intercalation and de-intercalation under the control of electric field with the SIC-FET device. Li intercalation leads to the formation of 1T$^{'}$-LiMoS$_2$ insulator, and the de-intercalation of Li ions changes the sample to 1T$^{'}$-MoS$_2$ superconductor. The transition between 1T$^{'}$-LiMoS$_2$ and 1T$^{'}$-MoS$_2$ are reversible, and the electronic properties and lattice structures can be well characterized by resistance and Raman spectra, respectively. Our methodology provides a generic way for regulating the structure and electronic properties in transition metal dichalcogenide thin flakes, which demonstrates a valuable technical approach for exploring novel physical properties in metastable lattice structures.

This work is supported by the National Key R$\&$D Program of the Ministry of Science and Technology of China (Grant Nos. 2017YFA0303001 and 2016YFA0300201), the National Natural Science Foundation of China (Grants Nos. 11534010 and 91422303), the Strategic Priority Research Program (B) of the Chinese Academy of Sciences (Grant No. XDB04040100) and the Key Research Program of Frontier Sciences CAS (Grant No. QYZDY-SSW-SLH021).

\end{document}